\documentclass[a4paper,twocolumn,11pt,accepted=2024-03-15]{quantumarticle}
\pdfoutput=1

\PassOptionsToPackage{compress}{natbib}
\usepackage[numbers]{natbib}
\usepackage{hyperref}

\usepackage{graphicx}% Include figure files
\usepackage{dcolumn}% Align table columns on decimal point
\usepackage{bm}% bold math

\usepackage[utf8]{inputenc}
\usepackage{amsmath,amsfonts,calc}
\usepackage{comment}   
\usepackage[normalem]{ulem}
\usepackage[usenames,dvipsnames]{color}
\usepackage{soul}
\usepackage{graphicx}
\usepackage{scalerel}
\usepackage{mathtools}
\usepackage{stackengine,wasysym}
\usepackage[free-standing-units=true]{siunitx}
\makeatletter
\newsavebox\myboxA
\newsavebox\myboxB
\newlength\mylenA
\makeatother

\newcommand{\ket}[1]{\left|#1\right>}

\begin{document}

\title{Effective versus Floquet theory for the Kerr parametric oscillator}
\author{Ignacio Garc\'ia-Mata}
\affiliation{Instituto de Investigaciones F\'isicas de Mar del Plata (IFIMAR), Facultad de Ciencias Exactas y Naturales,
Universidad Nacional de Mar del Plata \& CONICET, 7600 Mar del Plata, Argentina}

\author{Rodrigo G. Corti\~nas}
\affiliation{Department of Applied Physics and Physics, Yale University, New Haven, Connecticut 06520, USA}
\affiliation{Yale Quantum Institute, Yale University, New Haven, Connecticut 06520, USA}

\author{Xu Xiao}
\affiliation{Department of Applied Physics and Physics, Yale University, New Haven, Connecticut 06520, USA}

\author{Jorge Ch\'avez-Carlos}
\affiliation{Department of Physics, University of Connecticut, Storrs, Connecticut, USA}

\author{Victor S. Batista}
\affiliation{Department of Chemistry, Yale University, 
P.O. Box 208107, New Haven, Connecticut 06520-8107, USA}
\affiliation{Yale Quantum Institute, Yale University, New Haven, Connecticut 06520, USA}

\author{Lea F. Santos}
\affiliation{Department of Physics, University of Connecticut, Storrs, Connecticut, USA}

\author{Diego A. Wisniacki}
\affiliation{Departamento de F\'isica ``J. J. Giambiagi'' and IFIBA, FCEyN,
Universidad de Buenos Aires, 1428 Buenos Aires, Argentina}
\maketitle

\begin{abstract}
Parametric gates and processes engineered from the perspective of the static effective Hamiltonian of a driven system are central to quantum technology. However, the perturbative expansions used to derive static effective models may not be able to efficiently capture all the relevant physics of the original system. In this work, we investigate the conditions for the validity of the usual low-order static effective Hamiltonian used to describe a Kerr oscillator under a squeezing drive. This system is of fundamental and technological interest. In particular, it has been used to stabilize Schr\"odinger cat states, which have applications for quantum computing. We compare the states and energies of the effective static Hamiltonian with the exact Floquet states and quasi-energies of the driven system and determine the parameter regime where the two descriptions agree. Our work brings to light the physics that is left out by ordinary static effective treatments and that can be explored by state-of-the-art experiments. 
\end{abstract}

\section{\label{sec:introduction} Introduction}

Driven systems can present unexpected behaviors oftentimes without a static analog. A typical example is the Kapitza pendulum, where a rapidly driven rigid pendulum can stabilize against gravity  by developing a minimum of potential energy when pointing upward~\cite{Kapitza1951,landau1976}. In the quantum regime, this static effective potential was proposed as a way to generate an error-protected qubit~\cite{Venkatraman2021_PRL} and the electronic analog of this mechanical system was recently named ``Kapitzonium'' \cite{wang2023quantum}.
Another example are the Paul traps~\cite{Paul1990,goldman2014}, that use time-dependent electric fields to trap charged particles and therefore bypass Earnshaw's theorem, which states that a charge distribution cannot be stabilized in a stationary equilibrium configuration via its electrostatic interactions. By changing the electric field faster than the escaping rate of the particles, an average confining force can be created. This idea is at the basis of some atomic clocks and trapped-ion quantum technologies \cite{Wineland2013,bruzewicz2019trapped}. 

Static effective models are often used to study driven systems, because they provide analytical expressions and simplifications to the time-dependent problem.
 Many useful methods to achieve static effective Hamiltonians have been derived in the last century~\cite{Magnus1954,Fer1958,ErnstBook,haeberlenBook,Wilcox1967} (see \cite{Venkatraman2021_PRL} for a classification), but they are not without limitations.
In particular, in highly nonlinear systems, micromotion caused by the kicks of the rapidly changing potential can feedback into the dynamics and produce sizable effects associated with nonlinear mixing and amplifications~\cite{xiao2023diagrammatic}. These effects ultimately lead to regimes beyond the scopes of the static effective Hamiltonian. Alternatively, when the system is periodically driven, it can be studied numerically via Floquet theory. The advantage is that this method can be carried out with a minimal amount of approximations, yet it mostly remains a numerical treatment. 

In this work, we discuss to what extent the ordinary static effective theory and the numerical Floquet treatment agree.
We direct our attention to a dynamically rich system that is central to ongoing investigations: the Kerr parametric oscillator. It consists of a nonlinear oscillator subjected to a squeezing drive and boasts a storied history. It has served as an exemplary instance of a parametric oscillator \cite{marthaler2006,marthaler2007,dykman2012}, an amplifier \cite{Wustmann2013,krantz2016single,frattini2017}, a tool for stabilizing quantum information \cite{cochrane1999,Goto2016,Goto2019,Goto2021,puri2020}, a framework for quantum optical tuning \cite{wielinga1993}, and more recently, as a platform to study excited state phase transitions (ESQPTs) \cite{chavez2023} and tunneling~\cite{Prado2023}. The model has also been experimentally implemented with superconducting circuits, being employed to generate Schr\"odinger cat states~\cite{wang2019,grimm2020},  analyze tunneling~\cite{Venkat2022_delta,iyama2023observation}, and detect the exponential coalescence of pairs of energy levels as a function of a control parameter. This coalescence, named ``spectral kissing'' \cite{frattini2022squeezed}, is one of the features of ESQPTs.
 For the parameters values used, these experiments were well described by low-order static effective Hamiltonians. However, steady experimental progress and access to broader ranges of parameters prompt a deeper investigation beyond this regime. In this case, the full Floquet numerical analysis becomes useful.

Here, we quantify  the proximity of the driven and effective descriptions of the Kerr parametric oscillator  by comparing the Floquet quasienergies and Floquet states of the time-dependent Hamiltonian with the eigenenergies and eigenstates of the corresponding low-order effective Hamiltonian. This is done via numerical simulations covering experimentally relevant parameters regimes.
We find that deviations between the two approaches become larger as the parameters that control the nonlinear terms are increased. This is precisely the parameter region in which the experiments are moving toward. 

Understanding the parameter region where the effective low-order Hamiltonian, that has been used to describe recent experiments, is valid and reliable is of paramount importance for the design of quantum technologies, including qubits, gates, and circuits. Analyzing when low-order expansions may fail can point in the direction of new physics and new possibilities for applications in quantum information processing. 

The paper is structured as follows. In Sec.~\ref{sec:model}, we introduce the model system, the time-dependent Hamiltonian and its corresponding static effective approximation. In Sec.~\ref{sec:results}, we compare the eigenvalues and eigenvectors of the effective Hamiltonian with the quasienergies and Floquet states of the time-dependent system. This allows us to determine the experimentally accessible parameters regimes, where the two descriptions agree and where they are expected to diverge. Conclusions are presented in Sec.~\ref{sec:conclu} and additional results are given in the appendices.

\section{\label{sec:model} Model: Kerr parametric oscillator}

The Hamiltonian of the driven superconducting nonlinear oscillator -- the Kerr parametric oscillator -- that we analyze \cite{grimm2020,frattini2022squeezed,Venkat2022_delta} is analogous to a one-dimensional asymmetric driven quantum pendulum. In terms of dimensionless coordinates, the Hamiltonian is given by
\begin{align}
\begin{split}
    \frac{\hat{H}(t) }{\hbar} = \omega_o\hat a^\dagger \hat a &+  \frac{g_3}{3}(\hat a + \hat a^\dagger)^3 +\frac{g_4}{4}(\hat a + \hat a^\dagger)^4\\
    &- i\Omega_d (\hat a - \hat a^\dagger)\cos\omega_d t,
\end{split}
\label{eq:H1}
\end{align}
where $\omega_o$ is the bare frequency of the oscillator,  
$\hat{a}$ ($\hat{a}^{\dagger}$) is the bosonic annihilation (creation) operator satisfying the commutation relation $[\hat{a}, \hat{a}^{\dagger}] = 1$, the third and fourth order nonlinearities of the potential energy are $g_3, g_4\ll \omega_o$, and the drive is characterized by its strength $\Omega_d$ and its frequency $\omega_d$. Due to the cubic nonlinearity $g_3$, the Hamiltonian of the experimental system in Eq.~(\ref{eq:H1}) deviates from that for the simple Duffing oscillator.

The dimensionless operator $\hat{a} = \frac{1}{2}(\frac{\hat X}{ X_{\mathrm{zps}}} + i\frac{\hat P}{ P_{\mathrm{zps}}})$ is written in terms of the position-like $\hat X$ and the momentum-like $\hat P$ coordinates, the zero-point spread in the position-like degree of freedom of the oscillator $X_{\mathrm{zps}}=\sqrt{\hbar/(2M\omega_o)}$, and the zero-point spread in momentum $P_{\mathrm{zps}} = \hbar/(2X_{\mathrm{zps}})$, where $M$ is the effective mass of the oscillator (for the mapping to a superconducting quantum system, see \cite{Koch2004,girvin2014}). {Following the experiments~\cite{grimm2020,frattini2022squeezed,Venkat2022_delta}, we consider the condition to create parametric squeezing, that is, the system is driven at twice its bare oscillation frequency, $\omega_d\approx2\omega_o$. Under this condition, the system undergoes a period-doubling bifurcation~\cite{wielinga1993,cochrane1999,marthaler2006,marthaler2007,puri2017}, whose static effective description corresponds to a double-well system~\cite{marthaler2006,marthaler2007,dykman2012,chavez2023}.}

In Ref. \cite{frattini2022squeezed}, it was shown that the experiment carried out with the time-dependent Hamiltonian in Eq.~(\ref{eq:H1}) could be described by a low-order static effective Hamiltonian. To compute the effective Hamiltonian, two transformations must be applied to  Eq.~(\ref{eq:H1}). The first one is a displacement into the linear response of the oscillator, where the effective amplitude of the displacement is $\Pi \approx \frac{2\Omega_d}{3\omega_d}$. The second corresponds to a change into a rotating frame induced by $\frac{\omega_d}{2} \hat{a}^{\dagger} \hat{a}$, transforming Eq.~(\ref{eq:H1}) to 

\begin{align}
\label{eq:H}
\begin{split}
\frac{\hat{\mathcal{H}}(t)}{\hbar} =& -\delta \hat{a}^{\dagger} \hat{a} + \sum_{m = 3}^4 \frac{g_m}{m} (\hat{a} e^{-i \omega_d t/2}+ \\
&\ \hat{a}^\dagger e^{i \omega_d t/2}+ \Pi e^{-i \omega_d t} + \Pi^* e^{i \omega_d t})^m,
\end{split}
\end{align}
where $\delta = \frac{\omega_d}{2} - \omega_o \ll \omega_d $. This choice of frame brings the period doubling dynamics into focus. Note that the periodicity of Eq.~(\ref{eq:H}) is two times the periodicity of the drive (see Sec.B.I in \cite{marthaler2006} for a discussion).

The propagator over a period $T$ induced by Eq.~(\ref{eq:H}) is given by 
\begin{align}
    \hat U(T) = \mathcal{T}e^{\frac{-i}{\hbar}\int _0^{T} \hat{\mathcal{H}}(t)dt} = e^{\frac{i}{\hbar}\hat S(T)}e^{-\frac{i}{\hbar}\hat H_{\mathrm{eff}} T}e^{-\frac{i}{\hbar}\hat S(0)},
    \label{Eq:UandS}
\end{align}
where $\mathcal{T}$ is the time ordering operator, which does not appear on the right-hand side of the equation. The purpose of the operator $\hat{S}(t)=\hat{S}(t+T)$ is to generate a canonical transformation to a frame where the evolution is ruled by a time-independent Hamiltonian $\hat H_{\mathrm{eff}}$. This provides an important simplification. As discussed in detail in the next section, to compare the eigenstates of the effective Hamiltonian with the Floquet states, we have to take into account the unitary transformation, 
\begin{equation}
\hat{U}_S=e^{-\frac{i}{\hbar}\hat S},
\end{equation}
where $\hat S=\hat S(0)=\hat S(T)$.

So far no approximation has been made, but one can compute $\hat H_{\mathrm{eff}}$  and $\hat S$ perturbatively to arbitrary order using a mutually recursive formula developed in~\cite{Venkatraman2021_PRL}. Following the approach in \cite{Venkatraman2021_PRL,xiao2023diagrammatic}, one uses the zero-point spread of the oscillator, $X_{\mathrm{zps}}$, as the perturbation parameter and reaches the second-order effective Hamiltonian~\cite{frattini2022squeezed}
\begin{equation}
\label{eq:HKC}
\frac{\hat{H}^{(2)}_{\mathrm{eff}}}{\hbar} = \epsilon_2(\hat a^{\dagger2} + \hat a^{2}) - K\hat a^{\dagger2} \hat a^2 ,
\end{equation}
which conserves parity, $[\hat{H}^{(2)}_{\mathrm{eff}},e^{i\pi \hat{a}^\dagger \hat{a}}]=0$. The driving condition for the period-doubling bifurcation can now be better specified as $\omega_d = 2\omega_a$, where $\omega_a\approx\omega^{(2)}_a =\omega_o+3 g_4-20 g_3^2 / 3 \omega_o+$ $(6g_4 +9g_3^2/\omega_o)(2\Omega_d/3\omega_o)^2$ includes the Lamb and Stark shift to the bare frequency $\omega_o$. In Eq.~(\ref{eq:HKC}), the Kerr nonlinearity to leading order is $K \approx K^{(2)}= -\frac{3g_4}{2} +  \frac{10g_3^2}{3\omega_o}$ and $\epsilon_2 \approx \epsilon_2^{(2)} = g_3 \frac{2\Omega_d}{3\omega_o}$.  In these expressions, all nonlinear corrections are kept to order $X_{\mathrm{zps}}^2$ (see also Secs.~28 and 29 of \cite{landau1976} for explanations and derivations involving nonlinear oscillations and resonances). 

The experiments in \cite{Venkat2022_delta,frattini2022squeezed} have measured the spectrum up to the tenth excited state. For the parameters ranges that have been considered, the first ten eigenvalues of the Hamiltonian $\hat{H}^{(2)}_{\mathrm{eff}}$ in Eq.~(\ref{eq:HKC}) match the experimental results for the driven systems.

In what follows, we study the limits of applicability of the low-order static effective theory for a wide range of parameters of experimental interest. Various combinations of the native parameters $g_3$ and $g_4$ in Eq.~(\ref{eq:H1}) can yield the same emergent effective Kerr nonlinearity $K$, yet the effective Hamiltonian in Eq.~(\ref{eq:HKC}) is not equally good for all choices. This analysis is important for applications in quantum computing, where the Kerr nonlinearity needs to be much larger than the decoherence rate of the driven system \cite{puri2017,chamberland_building_2022,ruiz2022} to allow for the realization of several gates before the loss of coherence. 
The ability to engineer the correct static effective spectrum for the highly excited states of the driven nonlinear oscillator is also paramount for applications in quantum computing based on Kerr-cat qubits, because the autonomous error-protecting properties of the qubit encoded in the bifurcated oscillator depend strongly on the dynamics of the excited states~\cite{frattini2022squeezed,Venkat2022_delta,gautier_combined_2022,putterman_stabilizing_2022}.

Going beyond Eq.~(\ref{eq:HKC}),  we write down in appendix~\ref{appA} the fourth-order static effective Hamiltonian. In appendix~\ref{app:orders}, we extend the comparison between the time-dependent Hamiltonian in Eq.~(\ref{eq:H}) and the effective model in Eq.~(\ref{eq:HKC}), which is developed in the main text, to include also the fourth- and sixth-order effective Hamiltonians. The results are qualitatively  similar even though the high-order static effective expansion does not necessarily account for all nonlinear resonances caused by neglecting counter-rotating terms present in the time-dependent problem.

\section{\label{sec:results} Spectra of the Floquet and of the Effective Hamiltonian}

We start this section by comparing the solutions of the time-dependent system described by the periodically driven Hamiltonian $\hat{\mathcal{H}}(t)$ in Eq.~(\ref{eq:H})  with those of the effective Hamiltonian $\hat{H}^{(2)}_{\mathrm{eff}}$ in Eq.~(\ref{eq:HKC}). The time-dependent oscillator in Eq.~(\ref{eq:H})  relies on five parameters: $\omega_o, \;g_3,\; g_4,\; \omega_d$ and $\Omega_d$. From now on, we set $\omega_o=\hbar=1$.

We denote the eigenstates and eigenvalues of $\hat{H}^{(2)}_{\mathrm{eff}}$ as $|\psi_k \rangle$ and $E_k$, and its ground-state energy as $E_0$. The driven system is described by the Floquet states~\cite{shirley1965},
$$
|\Psi_{k} (t) \rangle = e^{-i \varepsilon_{k} t} |\phi_{k} (t) \rangle,
$$
where 
$|\phi_{k} (t) \rangle = |\phi_{k} (t+T) \rangle$
are the Floquet modes and $\varepsilon_{k}$ are the Floquet quasienergies. Due to the period doubling bifurcation, instead of using $T=T_d = 2 \pi/\omega_d$ as usually done in Floquet theory, we take the period $T$ to be twice the period of the drive $T_d$, that is $T=2T_d$.
Therefore,
our Floquet modes are 
the eigenstates of the time-evolution operator (Floquet operator) after two periods of the drive,
$$
\hat U(T) |\phi_{k} \rangle = e^{-i \varepsilon_{k} T}  |\phi_{k} \rangle ,
$$ 
and the quasienergies are obtained by diagonalizing $\hat U(T)$.  The quasienergies are uniquely defined modulo $\hbar \omega_d/2=2\pi \hbar /T$, that is, 
$$
\varepsilon_{k} \in [0, \hbar \omega_d/2].
$$

\subsection{Quasienergies vs Eigenvalues}

\begin{figure}[t]
\includegraphics[width=.45\textwidth]{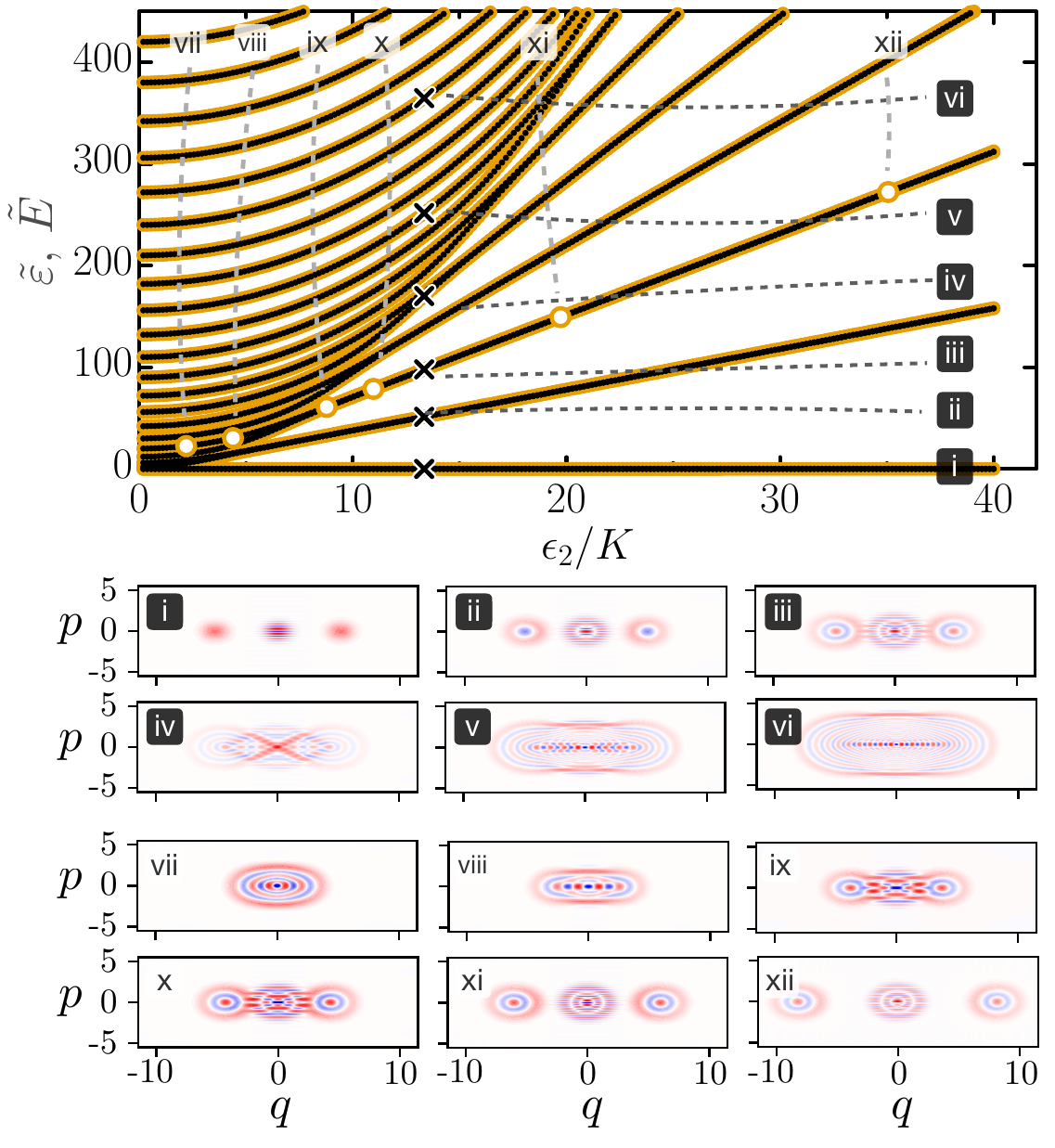} %{newnewfig1.pdf}
\caption{\label{fig:floqH4Wig}
Top panel: (Rescaled) Excitation energies $\tilde{E}=(E-E_0)/K$ (black lines) of $\hat{H}^{(2)}_{\mathrm{eff}}$ in Eq.~(\ref{eq:HKC}) and quasienergies $\tilde{\varepsilon}=[(\varepsilon-\varepsilon_0)\mod (\omega_d/2)]$ (orange) of $\hat{\mathcal{H}}(t)$ in Eq.~(\ref{eq:H}) as a function of the control parameter $\epsilon_2/K$. 
Panels (i)-(vi): Wigner functions of the Floquet states  corresponding to the quasienergies at the cross symbols \textbf{\textsf{x}} in the top panel for $\epsilon_2/K= 13$ and  $\tilde{\varepsilon}_k$ equal to (i) 0, (ii) 51.25, (iii) 97.9, (iv) 170.1 (v) 251.74, and  (vi) 364.76. States (i)-(iii) lie below the ESQPT and state (iv) is at the ESQPT critical energy for $\epsilon_2/K= 13$. Panels (vii)-(xii): Wigner functions of the Floquet states  corresponding to the circle symbols $\circ$ in the top panel for {quasienergy $\tilde{\varepsilon}_6$}  and $\epsilon_2/K$ equal to (vii) 2.19, (viii) 4.38, (ix) 8.76, (x) 10.96, (xi) 19.73, and (xii) 35.1. All panels: Basis size $N=200$, $g_3/\omega_o=0.00075$, $g_4/\omega_o=1.27\times 10^{-7}$,  and $K/\omega_o=1.685\times 10^{-6}$. In the Wigner functions, red corresponds to positive values and blue corresponds to negative values. }
\end{figure}

In the top panel of Fig.~\ref{fig:floqH4Wig}, we show the excitation energies $\tilde{E}=E-E_0$ (black lines) of $\hat{H}^{(2)}_{\mathrm{eff}}$  as a function of the control parameter $\epsilon_2/K$ and compare them with the quasienergies (orange lines) computed with respect to the ``ground-state quasienergy'' (see explanation below),  $\tilde{\varepsilon}=\varepsilon-\varepsilon_0$. To be able to compare $\tilde{E}$ and $ \tilde{\varepsilon}$ properly, we plot $\tilde{E}\equiv (E-E_0)/K$ and $\tilde{\varepsilon}=[(\varepsilon-\varepsilon_0)\mod (\omega_d/2)]/K$. For this figure, we chose the value of the nonlinearities to be $g_3/\omega_o=0.00075$, $g_4/\omega_o=1.27\times 10^{-7}$, and therefore $K/\omega_o=1.685\times 10^{-6}$. In what follows, we consider $g_4>0$, which allows for studying a wide range of values of $K$ with possible interesting applications~\cite{sivak_kerr-free_2019,frattini2017,Venkatraman2021_PRL}. The case $g_4<0$ is briefly explored in appendix B.  

%%%%%%%%%%%%%% FIGURE 1 %%%%%%%%%
%%%%%%%%%%%%%%%%%%%%%%%%%%%%

To determine {what we call} ground-state quasienergy $\varepsilon_0$, we use the method
developed in \cite{wisniacki2014universal}  to follow the eigenstates with definite localization properties.
We first determine the ground state {$\ket {\Psi_0}_{\epsilon_2/K=0}$ } of the  Hamiltonian in Eq.~(\ref{eq:H}) with $\Omega_d=0$.  We then turn on the drive with a small increment $\delta \Omega_d$ that results in an increase $\delta(\epsilon_2/K)$ of the control parameter. To determine the new Floquet ground state, {$\ket {\Psi_0}_{\delta(\epsilon_2/K)}$}, we search for the Floquet state that has the largest overlap with {$\ket {\Psi_0}_{\epsilon_2/K=0}$}. {We then use $\ket {\Psi_0}_{\delta(\epsilon_2/K)}$ to find  $\ket {\Psi_0}_{2\delta(\epsilon_2/K)}$.}
The procedure is repeated each time the drive amplitude is increased so that the Floquet ground state is recursively updated. 

For the small values of $g_3$ and $g_4$ considered in Fig.~\ref{fig:floqH4Wig}, $\hat{H}^{(2)}_{\mathrm{eff}}$ accurately describes the spectrum of $\hat{\mathcal{H}}(t)$. {As the control parameter $\epsilon_2/K$ increases, successive spacings between two adjacent levels, each level belonging to different parity sector, get exponentially small. This ``spectral kissing'' \cite{frattini2022squeezed} is a precursor of an ESQPT \cite{chavez2023}. The energy  where the levels merge together is the critical energy of the ESQPT, which grows quadratically with $\epsilon_2/K$.} The coalescence of the energies is accompanied by the clustering of the energy levels and the consequent divergence of the density of states at the ESQPT critical energy \cite{chavez2023}.

In Figs.~\ref{fig:floqH4Wig}(i)-(xii), we show the Wigner functions for the Floquet states. They are visually indistinguishable from the corresponding eigenstates of $H^{(2)}_{\rm eff}$ in this regime (see below).

The behavior of the states below and above the ESQPT is markedly different. Below the ESQPT, the system is represented by a double well and the  states exhibit a cat-like structure that can be used in quantum information processing \cite{mirrahimi2014dynamically}. This structure is revealed in Figs.~\ref{fig:floqH4Wig}(i)-(iii) by the Wigner functions of three Floquet states of $\hat{\mathcal{H}}(t)$ that show two  ellipses (one on the extreme left and the other on the extreme right of each panel) located at the minima of the effective double well, and in between them, centered at $q=0$, we see the interference fringes.  The  state  shown in Fig.~\ref{fig:floqH4Wig}(iv) is at the ESQPT critical energy, being highly localized at the Fock state $|0\rangle$, which translates into a state concentrated at the origin of the classical phase space, where there is an unstable hyperbolic point~\cite{chavez2023}. The Wigner function of this state in Fig.~\ref{fig:floqH4Wig}(iv) is visibly localized along the separatrix (classical stable and unstable manifolds).  Above the ESQPT,  as the energy increases, the Floquet states approach the eigenstates of a harmonic potential, as seen in Figs.~\ref{fig:floqH4Wig}(v)-(vi). 

In Figs.~\ref{fig:floqH4Wig}(vii)-(xii), we show the Wigner functions of the Floquet state with quasienergy {$\tilde{\varepsilon}_6$ } for different values of $\epsilon_2/K$. They are marked with circles in the top panel of Fig.~\ref{fig:floqH4Wig}. For the value of $\epsilon_2/K$ in Fig.~\ref{fig:floqH4Wig}(vii), 
this  Floquet state is above the ESQPT and resembles an eigenstate of a harmonic potential, but for the values of $\epsilon_2/K$ in Figs.~\ref{fig:floqH4Wig}(x)-(xii), the state is below the ESQPT and turns into a cat-state like.

%%%%%%%% FIGURE 2 %%%%%%%%%%%%%%%%%% 
\begin{figure}
\includegraphics[width=.48\textwidth]{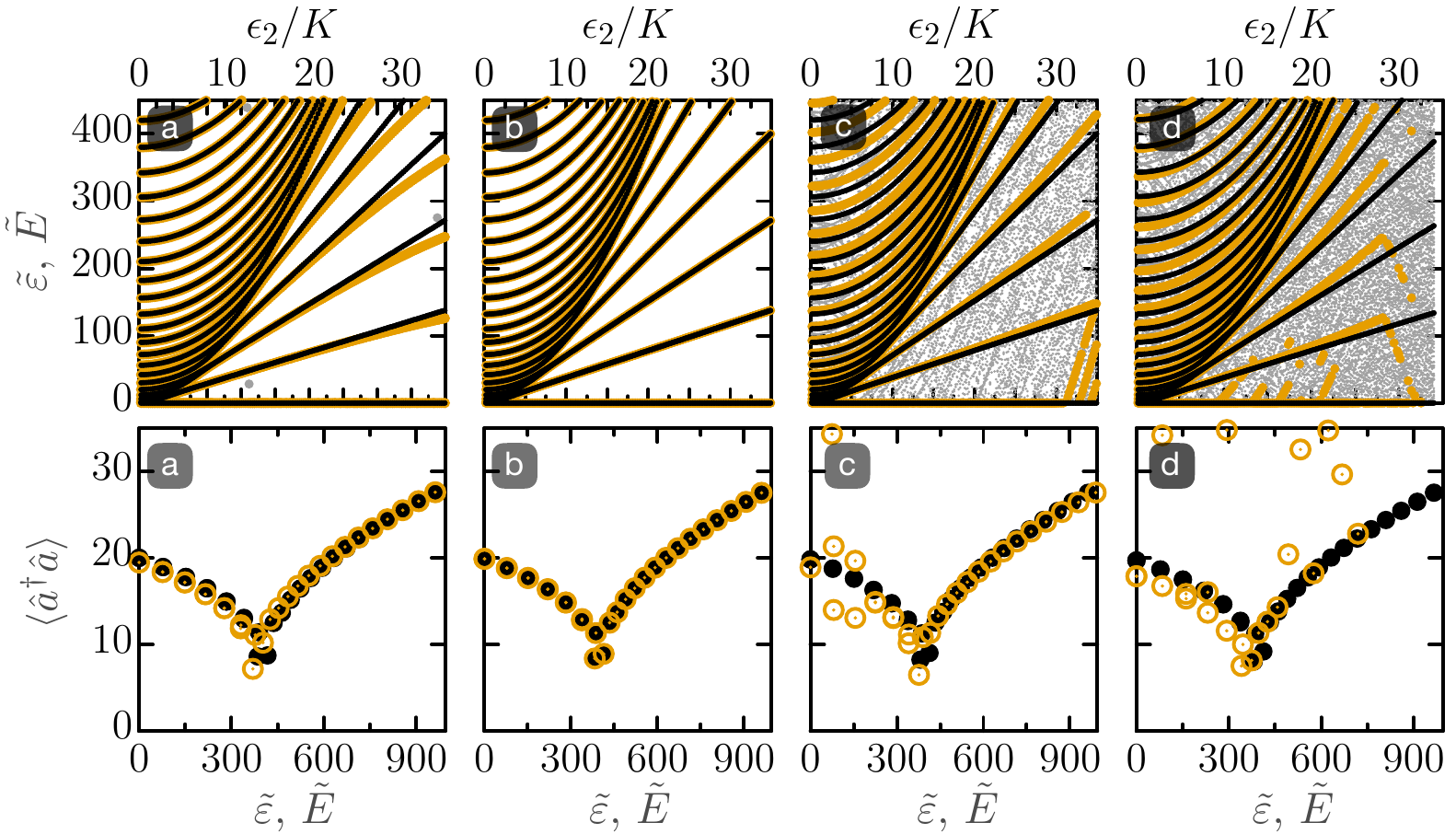} %%%{newfig2_abcd.png}
\caption{\label{fig:com} 
Top row (a)-(d): Excitation energies $\tilde{E}=E-E_0$ (black lines) of $\hat{H}^{(2)}_{\mathrm{eff}}$ in Eq.~(\ref{eq:HKC}) and quasienergies $\tilde{\varepsilon}=\varepsilon-\varepsilon_0$ (orange) of $\hat{\mathcal{H}}(t)$ in Eq.~(\ref{eq:H}) as a function of the control parameter $\epsilon_2/K$. The gray dots in (c)-(d) correspond to Floquet states that have, respectively, $\langle \phi|\hat a^\dag \hat a|\phi\rangle>28 ,\, 20$. 
Bottom row (a)-(d): $\langle E|\hat a^\dag \hat a|E\rangle$ (black circles) [$\langle \phi|\hat a^\dag \hat a|\phi\rangle$ (orange circles)] as a function of  $\tilde{E}$ {[$\tilde{\varepsilon}$]} for $\epsilon_2/K=20$. The nonlinearity strengths $(g_3,g_4)$ are $(2\times 10^{-5},8\times 10^{-6})$ for (a); $(0.00075,1.27\times 10^{-7})$ for (b);  $(0.015,10^{-7})$ for (c); and $(0.02,10^{-7})$ for (d). Basis size is $N=200$}
\end{figure}
%%%%%%%%%%%%%%%%%%%%%%%%%%%%%%%%%% 

In Fig.~\ref{fig:com}, we continue our comparison of the quasienergies (orange lines) of the driven Hamiltonian in Eq.~(\ref{eq:H})  with the eigenenergies (black lines) of the effective Hamiltonian $\hat{H}^{(2)}_{\mathrm{eff}}$ in Eq.~(\ref{eq:HKC}). {However, for a good visual comparison between $\tilde{\varepsilon}$ and $\tilde{E}$, we distinguish the quasienergies for the states associated with a small mean number of photons $\langle \phi|\hat{a}^\dagger \hat{a}|\phi\rangle$ (orange) -- smaller than some maximum value that is set differently depending on $g_3$ and $g_4$ --  with those that have a large mean number of photons (gray)}. The values of $g_3$ and $g_4$ considered lie within reach of current experimental setups. In the Fig.~\ref{fig:com}(a) of the top row, the choices of $g_3$ and $g_4$ lead to $K<0$, while the Figs.~\ref{fig:com}(b)-(d) of the top row have $K>0$. Notice that the Fig.~\ref{fig:com}(b) of the top row is the same as Fig.~\ref{fig:floqH4Wig} and it exhibits almost perfect coincidence between $\tilde{E}$ and $\tilde{\varepsilon}$. This panel is shown here again for a good comparison with the other cases. 

Spectral kissing is still observed in the Fig.~\ref{fig:com}(a) of the top row, {despite the negative value of the Kerr amplitude $K$}, but the agreement between $\tilde{E}$ and $\tilde{\varepsilon}$ deteriorates for {the} larger values of $\epsilon_2/K$ {in the Figs.~\ref{fig:com}(c)-(d) of the top row. For these two panels,} in addition to the reduced agreement between the spectrum of $\hat{H}^{(2)}_{\mathrm{eff}}$ (black lines) and part of the quasienergies of $\hat{\mathcal{H}}(t)$ (orange points), we also show with gray points the quasienergies that have no relationship with $\tilde{E}$.  These are the quasienergies that are folded to the first Brillouin zone. Since the quasienergies are defined modulo $\hbar \omega_d/2$, after folding them to the first Brillouin zone, they get clustered, as seen with those gray points. This issue becomes more evident in the Figs.~\ref{fig:com}(c)-(d) of the top row, where $g_3$ is larger than in the Fig.~\ref{fig:com}(b) of the top row. 

%%%%%%%%%%% FIGURE 3 %%%%%%%%%%%%%%% 
\begin{figure*}
\includegraphics[width=1\textwidth]{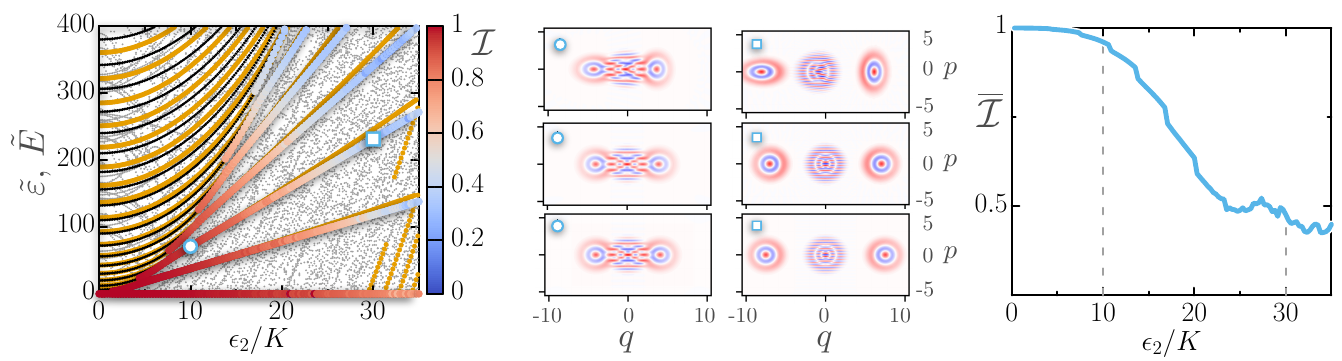}  %%{fig3_landscape.png}
\caption{\label{fig:2panels} 
Left panel: Same as the Fig.~\ref{fig:com}(c) in {the top of row}, but showing also {with a gradient of colors} the inverse participation ratio, ${\cal I}$, of each eigenstate below the ESQPT. The circle and the square symbols mark the state at $\epsilon_2/K=10$ and $\epsilon_2/K=30$, respectively.
Middle of the figure: Left (Right) column gives the Wigner functions for the state marked with a circle (square) on the left panel {for the spectrum}; top panel in the column corresponds to the Floquet state, middle panel in the column corresponds to the Floquet state after the $\hat{U}_S$ transformation, and bottom panel in the column corresponds to the eigenstate of $\hat{H}_{\rm eff}^{(2)}$. 
Right panel: Average IPR, $\overline{\cal I}$, for the eigenstates below the ESQPT {using Eqs.~(\ref{eq:IK})-(\ref{eq:xi})}. All panels: Basis size $N=200$ , $g_3/\omega_o=0.015$ and $ g_4/\omega_o=10^{-7}$.
} 
\end{figure*}
%%%%%%%%%%%%%%%%%%%%%%%%%%%%%%%%%% 

In addition to the spectral kissing and the divergence of the density of states, the presence of an ESQPT is also characterized by a discontinuity in some observables at the ESQPT energy~\cite{chavez2023,santos2016excited}. This can be seen in the bottom row of Fig.~\ref{fig:com}, where we show the average number of photons for each excited state as a function of the rescaled excitation energy (quasienergy) $\tilde{E}$ ($\tilde{\varepsilon}$) for $\epsilon_2/K=20$ for the same nonlinearities used in the top row of Fig.~\ref{fig:com}. The sudden drop in the values of $\langle \phi|\hat{a}^\dagger \hat{a}|\phi\rangle$ at the ESQPT critical energy is visible in Figs.~\ref{fig:com}(a)-(b) of the bottom row. This becomes less evident in the Fig.~\ref{fig:com}(d) of the bottom row, reflecting the weaker agreement between eigenvalues and quasienergies in the Fig.~\ref{fig:com}(d) of the top row.

\subsection{Floquet States vs Eigenstates}

A deeper comparison between the driven system and its corresponding effective model can be achieved through the analysis of the structure of the Floquet states of $\hat{\mathcal{H}}(t)$ and the eigenstates of $\hat{H}^{(2)}_{\mathrm{eff}}$.
Due to the importance of the cat-like states for quantum information processing, we concentrate our analysis on the states that lie below the ESQPT. 

To quantify the proximity of an eigenstate $|E_k\rangle$ of the effective Hamiltonian to a Floquet state, we use the  
inverse participation ratio (IPR), which is a measure of the level of delocalization of quantum states (see \cite{RevModPhys.80.1355} and references therein). Here, we define the IPR of state $|E_k \rangle$ as
$I_k = \sum_j |a_j|^4$,
where $a_j$ are the coefficients given by the expansion of the eigenstate in the basis of the Floquet modes, that is, $|E_k\rangle = \sum_j a_j|\phi_j\rangle$. The IPR ranges from $1/N$ for a completely delocalized state in the given basis ($N$ is the number of basis states) to 1 for a state that is completely localized in a single basis state. 

There is, however, one further aspect that needs to be taken into account. According to Eq.~(\ref{Eq:UandS}), the Floquet modes $\ket{\phi_j}$ and the effective Hamiltonian eigenstates  $\ket{E_k}$ live in different reference frames separated by $\hat S$. To be able to compare them in the same frame, we need to compute the IPR defined as 
\begin{equation}
\label{eq:IK}
{\cal I}_k = \sum_j |\langle \phi_j|\hat{U}_{S}|E_k\rangle|^4.
\end{equation}
In addition to ${\cal I}_k$, the other quantity that we use as a figure of merit is the  average 
\begin{equation}
\label{eq:xi}
\overline{{\cal I}}=\frac{1}{n_b}\sum_{k=1}^{n_b} {\cal I}_k ,
\end{equation}
where $n_b$  $\approx 2\frac{\epsilon_2}{\pi K}$ is the number of states of the effective Hamiltonian below the energy of the ESQPT ($E_k\leq  \epsilon_2^2/K$) obtained in \cite{chavez2023,puri2017}.

In the left panel of Fig.~\ref{fig:2panels}, we show $\overline{{\cal I}}$ (colored lines) as a function of $\epsilon_2/K$ for each eigenstate of $\hat{H}^{(2)}_{\rm eff}$ with energy below the ESQPT. The parameters are the same as in Fig.~\ref{fig:com}(c).  One sees that as $\tilde{E}$ (black lines) and $\tilde{\varepsilon}$ (orange points) distant themselves in the Fig.~\ref{fig:com}(c) of the top row, ${\cal I}$ decreases significantly in the left panel of Fig.~\ref{fig:2panels}. This separation between the spectrum of the driven system and of the effective model is more visible for larger values of the control parameter $\epsilon_2/K$ and for larger energies. In this region, higher orders of the effective Hamiltonian are needed to get better agreement.

In the two columns of panels in the middle of Fig.~\ref{fig:2panels}, we have a closer look at the structure of two selected states, which are marked with a circle (at $\epsilon_2/K=10$) and a square (at $\epsilon_2/K=30$) in the left panel of Fig.~\ref{fig:2panels}. The left column of panels is for Wigner function of the state at $\epsilon_2/K=10$ and the right column for the state at  $\epsilon_2/K=30$. The top panels of the columns give the Wigner function for the Floquet state, the middle panels are for the Floquet state after the $\hat{U}_S^\dagger$ transformation, and the bottom panels for the eigenstate of $\hat{H}_{\rm eff}^{(2)}$. After the $\hat{U}_S^\dagger$ transformation, the Floquet states approach the eigenstates of $\hat{H}_{\rm eff}^{(2)}$. 

%%%%%%%%%%%%%%%%% FIG 4 %%%%%%%%%%%%%%%%
\begin{figure}[t]
\includegraphics[width=.9\linewidth]
{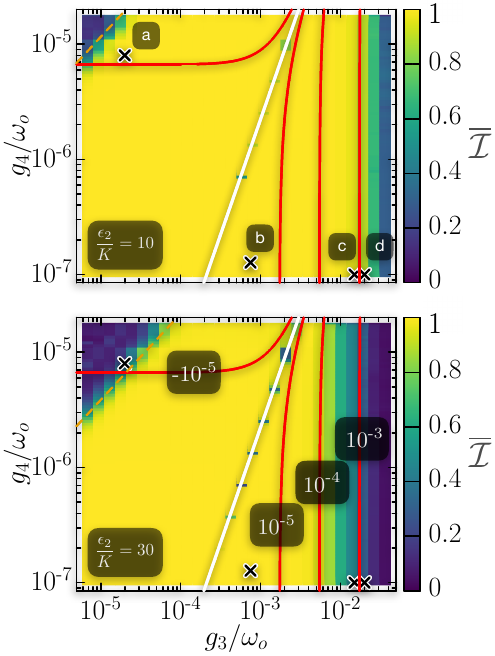}  %%{newfigIPRg3pos.png} 
\caption{\label{fig:2panels-2} 
Average IPR,  {$\overline{{\cal I}}$}, computed for the eigenstates of $H_{\rm eff}^{(2)}$ in the Floquet basis {using Eqs.~(\ref{eq:IK})-(\ref{eq:xi})} as a function of $g_3/\omega_o$ and $g_4/\omega_o$,  for $\epsilon_2/K=10$ (top panel) and $\epsilon_2/K=30$ (bottom panel). The white line in both panels marks $K=0$ and the basis size is $N=150$. The crosses in both panels (labelled a-d in the top panel) mark the points corresponding to the values of $g_3$ and $g_4$ used, respectively, in Figs.~\ref{fig:com}(a)-(d). The red lines in both panels are constant $K$ lines corresponding to $K/\omega_o=-10^{-5},\,\,10^{5},\,10^{4},\, 10^{-3}$, as indicated in the bottom panel.  The dashed orange line in both panels corresponds to the empirically determined boundary $g_4=a g_3^{3/4}/(\epsilon_2/K)$ with $a=0.65$.}  
\end{figure}
%%%%%%%%%%%%%%%%%%%%%%%%%%%%%%%%%%%%%%%%%%%%%%
%
%%%%%%%%%%%%%%%%% FIG 5 %%%%%%%%%%%%%%%%
\begin{figure}[t]
\includegraphics[width=.9\linewidth]{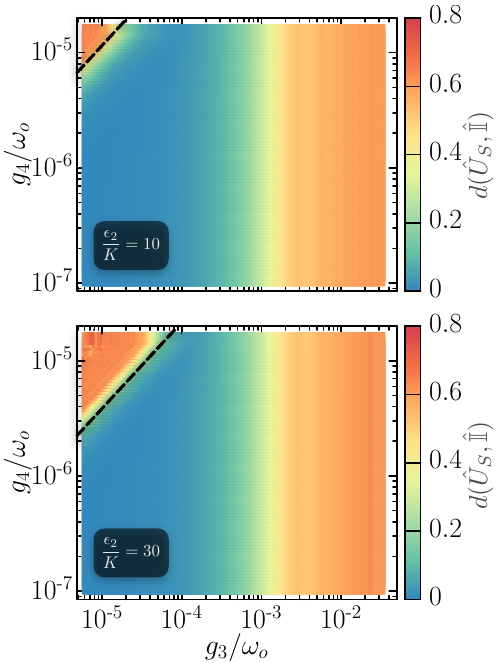}  %%{fig_diffUS.pdf} 
\caption{\label{fig:diffUS} Trace distance between $\hat{U}_S=e^{-i \hat S}$  and the identity operator $\hat{\mathbb{I}}$ as a function of $g_3/\omega_o$ and $g_4/\omega_o$ for $\epsilon_2/K=10$ (top panel) and $\epsilon_2/K=30$ (bottom panel). $N=100$. The dashed orange line in both panels corresponds to the empirically determined boundary $g_4=a g_3^{3/4}/(\epsilon_2/K)$ with $a=0.65$.}
\end{figure}
%%%%%%%%%%%%%%%%%%%%%%%%%%%%%%%%%

In the right panel of Fig.~\ref{fig:2panels}, we show the average $\overline{\cal I}$ for the states below the ESQPT as a function of $\epsilon_2/K$. For small  $\epsilon_2/K$, reflecting the left panel of Fig.~\ref{fig:2panels}, $\overline{\cal I}\approx 1$. As $\epsilon_2/K$ increases, the Floquet states and eigenstates drift apart,  more excited states fall under the well (that is,  $n_b$ increases), and $\overline{\cal I}$ gradually decreases. 

To gain further insight into the structures of the states and their dependence on the nonlinearities, we perform in Fig. \ref{fig:2panels-2} a 
systematic study of the mean value of the IPR,  defined in Eq.~(\ref{eq:xi}), as a function of $g_3$ and $g_4$. In the top panel of Fig. \ref{fig:2panels-2}, $\epsilon_2/K=10$, in the bottom panel, $\epsilon_2/K=30$, and the thick white line in both panels marks the values of $g_3$ and $g_4$ that lead to $K=0$. 
We observe that there is a yellow region around $K=0$, where ${\cal I}\approx 1$, which indicates that the eigenstates of $\hat H_{\rm eff}^{(2)}$ describe very well the Floquet states of the driven Hamiltonian in Eq.~(\ref{eq:H}). 

Comparing the top and bottom panels of Fig.~\ref{fig:2panels-2}, we see that the yellow region, {which indicates} excellent agreement between eigenstates and Floquet states, decreases as $\epsilon_2/K$ increases. This is evident for small values of $g_3$, where the yellow region is bounded  by a power-law curve (dashed orange line) empirically found to be approximately $g_4\sim g_3^{3/4}/(\epsilon_2/K)$, and for large values of $g_3$, where the region of low values of ${\cal I}$ (blue) grows to the left.

The cross symbols marked with the letters a, b, c, and d  in Fig.~\ref{fig:2panels-2} correspond to the values of $g_3$ and $g_4$ used, respectively, in Figs.~\ref{fig:com}(a), (b), (c), and (d). One sees that even though the average IPR in point a, ${\cal I}\approx 1$, indicates significant localization for $\epsilon_2/K=10$ in the top panel of Fig.~\ref{fig:2panels-2}, the same point shows delocalization (${\cal I}\approx 0$) for $\epsilon_2/K=30$ in the bottom panel. This behavior matches what is observed for the spectrum in the Fig.~\ref{fig:com}(a) of the top row. The correspondence with the top row of Fig.~\ref{fig:com} holds for the other three points b, c, and d. Points c and d, in particular, have large values of the $g_3$-nonlinearity, where the agreement between eigenstates and Floquet states are expected to break down for large $\epsilon_2/K$, as indeed verified in the bottom panel of Fig.~\ref{fig:2panels-2}. 

The red lines in Fig.~\ref{fig:2panels-2} indicate constant values of $K/\omega_o$. They serve as reference for experimental designs, so that one can know for which values of $g_{3}/\omega_o$ and $g_{4}/\omega_o$,  the constant $K/\omega_o$ is inside a region of agreement between Floquet states and eigenstates.

In the appendix~\ref{app:negg4},  we show a figure equivalent to Fig.~\ref{fig:2panels-2}, but for $g_4<0$. This is done, because it can be important for current and future experiments.  In this case, one cannot reach  $K=0$, but we verify that the results are very similar to those found for $g_4>0$ in Fig.~\ref{fig:2panels-2}.

The message conveyed by Fig.~\ref{fig:2panels-2} is rich and subtle. One could straightforwardly infer that inside the yellow region, where $\bar{\cal{I}}\approx 1$ , the static effective theory describes well the Floquet system. But this is true provided the transformation $\hat{U}_S$ is taken into account. To better understand this point, we show in Fig.~\ref{fig:diffUS} the distance between $\hat{U}_S$ and the identity operator $\hat{\mathbb{I}}$, defined as 
\begin{equation}
    d(\hat{U}_S,\hat{\mathbb{I}})=\frac{1}{2N}
    \|\hat{U}_S-\hat{\mathbb{I}}\|,
\end{equation}
where $\|\cdot\|$ is the trace norm.

In the regime of parameters where $d(\hat{U}_S,\hat{\mathbb{I}})\ll1$ (Fig.~\ref{fig:diffUS}) we find that all the static effective theory is contained within the effective Hamiltonian: the complicated frame transformation $\hat{U}_S$ away from the (trivially displaced and rotating) lab frame is negligibly small. In this regime, a large overlap between eigenstates and Floquet states ($\bar{\cal{I}}\approx 1$, Fig.~\ref{fig:2panels-2}) is guaranteed.

{Overall, whenever the agreement between the eigenstates of $H_{\rm eff}^{(2)}$ and the Floquet states is poor (blue region in Fig.~\ref{fig:2panels-2}), $d(\hat{U}_S,\hat{\mathbb{I}})$ is large in Fig.~\ref{fig:diffUS} (orange region). However, there are regions in which $d(\hat{U}_S,\hat{\mathbb{I}})$ is large, while Fig.~\ref{fig:2panels-2} still suggests good eigenstate-Floquet-state agreement. These cases indicate that a}
sizable $d(\hat{U}_S,\hat{\mathbb{I}})$ may still allow for a large overlap, provided the frame transformation modifying the eigenstates of the effective Hamiltonian is taken into account when comparing them to the Floquet states. The relevant states in the lab frame may look very different from the eigenstates of the static effective Hamiltonian. That is, Fig.~\ref{fig:diffUS} measures the distance between the reference frame in which Eq.~(\ref{eq:H}) {describes} the system and the frame generated by $\hat S$ where the static effective description is valid. 

We then remark that {the right panel of} Fig.~\ref{fig:2panels-2} measures the accuracy of the full static effective theory {taking $\hat{U}_S$ into account}, while Fig.~\ref{fig:diffUS} measures the accuracy of the usual static effective Hamiltonian treatment, which assumes that Eq.~(\ref{eq:H}) and the static effective Hamiltonian describe the system in the same frame \cite{grimm2020,frattini2022squeezed,Venkat2022_delta,iyama2023observation,wang2019}.

{We finish this section with a discussion about open systems. Whether the environment corresponds to 
a measurement device, a coherent control pulse, or a heat reservoir, one expects that the operator $\hat{U}_S$, which is iteratively constructed to produce a static effective description of the system alone, will modify the coupling to the environment.} This could lead to measurement infidelity, control anomalies, or exotic forms of nonlinear-driven dissipation \cite{dykman1975spectral}.  Presently, this nonlinear-driven dissipation is under investigation as a potential reason for discrepancies observed between open quantum system experiments and theoretical models  \cite{venkatraman2022Lindbladian}. Our analysis {calls attention to the importance of including $\hat{U}_S$ in the studies of open systems as well.}

\section{\label{sec:conclu} Final remarks}
We analyzed the conditions under which the time-dependent Hamiltonian {that describes} a driven superconducting circuit can be approximated by low-order effective time-independent Hamiltonians. 
Our focus was on the part of the spectrum below the ESQPT, where the states exhibit cat-like features that are relevant for quantum computing and quantum information science.

We found that there exists a well-defined region of values of the nonlinearity parameters $g_3$ and $g_4$, where the eigenvalues and the eigenstates of the effective Hamiltonian $\hat H _{\textrm{eff}}^{(2)}$ describe correctly the quasienergies and the Floquet states of the time-dependent system. However, in the limit of large values of  $g_3$ or  $g_4$, the correspondence breaks down. The phase diagram of the nonlinearities $g_3$ and $g_4$ that we provided for the analysis of coincidence between the effective and driven models has practical implications for the design of Kerr parametric oscillators and other parametric processes that tend to be overlooked. 

An important conclusion of our study is that the effective Hamiltonian suffices for the comparison between the quasienergies of the driven system and the eigenvalues of the static effective description, but the comparison between states requires also the analysis of the unitary transformation $\hat{U}_S$. The static effective theory includes both the effective Hamiltonian and $\hat{U}_S$. Without the latter, one may infer the failure of a given order of the effective description for parameter values, where it may actually still hold. 

In the appendix~\ref{app:orders},  we showed that the agreement between the driven and effective model can hold for larger values of the nonlinearities if one increases the perturbation order. This raises the question of whether one could expect exact agreement between the driven and static descriptions for an infinite order. The answer is negative, because for large nonlinearities and strong drive, the driven system can develop chaos, as shown in~\cite{ChavezPrep}. {When chaos sets in,} there is no perturbation order that can lead to agreement with the static effective Hamiltonian, which is necessarily integrable. Chaos can produce the collapse of the ESQPT \cite{GmataCriticality2021}, while the static effective description is integrable by construction. 

{The impact of our work is twofold. It is of interest to those aiming at generating  Schr\"odinger cat states and analyzing ESQPT. This group can use our ($g_3$, $g_4$) map to guarantee that they stay within the region where the static effective describes well the driven system. But our analysis is also of interest to those who want to explore new physics beyond the validity of static effective models. In particular, as nonlinearities are pushed to larger values, chaos will eventually set in. This reveals a new application for driven Kerr oscillators, that of devices to explore the effects of chaos, as explained in Ref.~\cite{ChavezPrep}.}

\begin{acknowledgments}
 %The authors thank X. YZ for interesting discussions. 
This work was supported by the
NSF CCI grant (Award Number 2124511). D.A.W and I.G.-M. received support from CONICET (Grant No.~PIP 11220200100568CO), UBACyT (Grant  No.~ 20020170100234BA)  and  ANCyPT (Grants No.~PICT-2020-SERIEA-00740 and PICT-2020-SERIEA-01082). I.G.-M. received support from CNRS (France) through the International Research Project (IRP) ``Complex Quantum Systems'' (CoQSys).

\end{acknowledgments}

\appendix

%%%%%%%%%%%%%%%%%%%
\section{Fourth-order effective Hamiltonian}
\label{appA}
At fourth order, the effective Hamiltonian includes  a four-photon drive and the first non-squeezing drive term resulting in 
\begin{eqnarray}
\frac{\hat{H}_{\mathrm{eff}}^{(4)}  }{\hbar} = &-&\Delta^{(4)} \hat{a}^{\dagger} \hat{a} - K^{(4)} \hat{a}^{\dagger 2} \hat{a}^2  - \lambda^{(4)} \hat{a}^{\dagger 3} \hat{a}^3 \nonumber \\ &+&  \epsilon^4_{(4)} \hat{a}^{\dagger 4} + \epsilon_{(4)}^* \hat{a}^{4} ,
\label{eq:H-4}
\end{eqnarray}
where 
\begin{align}
\label{eq:order4terms0}
\Delta^{(4)} &=  \sum_{k = 0, 1, 2} \Delta^{(4)}_{[k]}  |\Pi|^{2k}, \\
\label{eq:order4terms1}
K^{(4)} &= \sum_{k = 0, 1} K^{(4)}_{[k]}  |\Pi|^{2k},    
\end{align}
 and displayed below, are the analytical expressions for all the coefficients of Eq.~(\ref{eq:H-4}):
\begin{align}
\label{eq:order4terms}
\begin{split}
-\Delta^{(4)}_{[0]} &=   \frac{9 g_4^2}{\omega_a}    + 47 \frac{g_3^2 g_4}{\omega_a^2}    - \frac{6269}{324} \frac{g_3^4}{\omega_a^3}          \\
-\Delta^{(4)}_{[1]} &=  \frac{54}{5}  \frac{g_4^2}{\omega_a}      + \frac{671}{10}  \frac{g_3^2 g_4}{\omega_a^2 }      + \frac{113}{360}  \frac{g_3^4}{\omega_a^3}        \\
-\Delta^{(4)}_{[2]} &=  -    \frac{9}{2} \frac{g_4^2}{\omega_a}       + \frac{15113}{600} \frac{g_3^2 g_4}{\omega_a^2}      -\frac{297947}{32400} \frac{g_3^4}{\omega_a^3}     \\
-K^{(4)}_{[0]} &=  \frac{153}{16} \frac{g_4^2}{\omega_a}     + \frac{225}{4} \frac{g_3^2 g_4}{\omega_a^2}     +  \frac{805}{36} \frac{g_3^4}{\omega_a^3}    \\
-K^{(4)}_{[1]} &=    \frac{27}{5} \frac{g_4^2}{\omega_a}     + \frac{671}{20} \frac{g_3^2 g_4}{\omega_a^2}      +  \frac{113}{720}\frac{g_3^4}{ \omega_a^3}     \\
-\lambda^{(4)} &=  \frac{17}{8} \frac{g_4^2}{\omega_a} +  \frac{25}{2} \frac{g_3^2 g_4}{ \omega_a^2}    + \frac{805}{162} \frac{g_3^4}{\omega_a^3}     \\
\epsilon_4^{(4)} &=\left(  \frac{33}{8} \frac{g_4^2}{\omega_a}    -\frac{101}{96} \frac{g_3^2 g_4}{\omega_a}     -\frac{2009}{1296} \frac{g_3^4}{\omega_a^3}     \right)\Pi^2.
\end{split}
\end{align}

We observe that a necessary condition for the validity of the order-two static effective theory is that the next corrections be negligible:
$$\lambda^{(4)} \ll (K^{(2)}-K_{[0]}^{(4)})/n.$$
This condition clearly depends on both $g_3$ and $g_4$, even in absence of the drive. Note that for any given value of the nonlinearities, for a sufficiently large Fock number $n$, the approximation will fail.
%%%%%%%%%%%%%%%%%%%%%%%%%%%
\section{Convergence of the static effective theory}
\label{app:orders}

In the main text, we focused on the static effective description of the Kerr parametric oscillator to the first nontrivial order, which corresponds to the second-order effective Hamiltonian $\hat{H}^{(2)}_{\rm eff}$. In this appendix, we evaluate the convergence of the high order static effective theory when applied to the Kerr parametric oscillator.
This is done by comparing the fourth-order (see appendix~\ref{appA}) and the sixth-order static effective Hamiltonians, { $\hat{H}^{(4)}_{\rm eff}$ and  $\hat{H}^{(6)}_{\rm eff}$}, with the driven Hamiltonian in Eq.~(\ref{eq:H}).

%%%%%%%%% FIGURE 6 %%%%%%%%%%%%
\begin{figure}
\includegraphics[width=.9\linewidth]{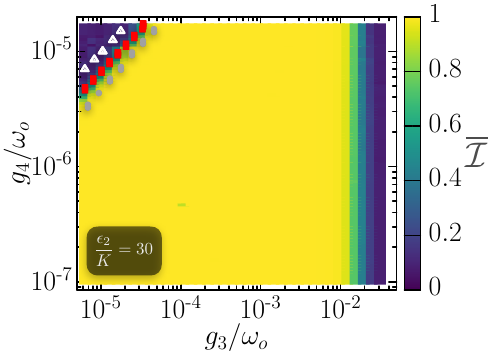} %%{figH4.pdf}
\caption{\label{fig:IPRH4} 
Average IPR, {$ \overline{{\cal I}}$ }, as a function of $g_3/\omega_o$ and $g_4/\omega_o$ computed using $H_{\rm eff}^{(4)}$ for $\epsilon_2/K=30$. The red squares mark the boundary between delocalization and localization set to {$ \overline{{\cal I}}\approx0.5$}. The gray circles and white triangles mark the same boundary for $H_{\rm eff}^{(2)}$ (as shown in Fig.~\ref{fig:2panels-2}) and $H_{\rm eff}^{(6)}$, respectively. Basis size is $N=150$.
}
\end{figure}
%%%%%%%%%%

Motivated by experimental developments in the field of quantum circuits, two methods to go beyond this first-order theory in a systematic manner were derived {in} \cite{Venkatraman2021_PRL, xiao2023diagrammatic}. These approaches, {which are} useful to explain experimental data and numerical simulations \cite{Venkatraman2021_PRL, xiao2023diagrammatic}, requires a symbolic computer program to carry out the analytical calculation, since the number of terms is far too great to write down by hand. 
%%%%%%%%%%% FIGURE 7 %%%%%%%%%%%%%%%%%%%
\begin{figure}[t]
\includegraphics[width=.9\linewidth]{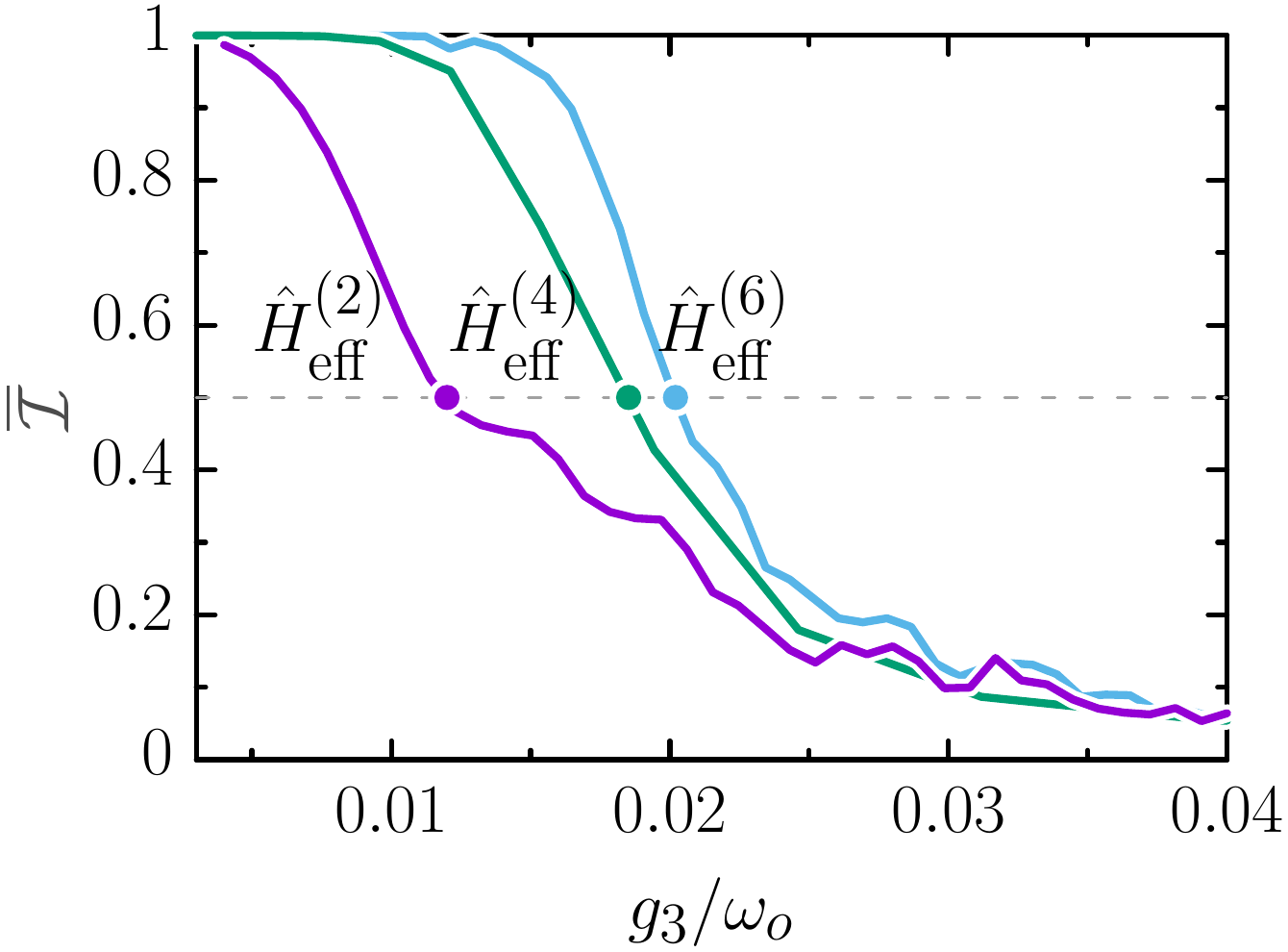} %%{newfig8.pdf}
\caption{\label{fig:order} 
Average IPR, {$\overline{{\cal I}}$}, as a function of $g_3/\omega_o$  computed using $\hat{H}_{\rm eff}^{(2)}$, $\hat{H}_{\rm eff}^{(4)}$, and $\hat{H}_{\rm eff}^{(6)}$, as indicated, for $N=100$, $\epsilon_2/K=30$ and {$g_4/\omega_o=10^{-7}$}. 
}
\end{figure}
%%%%%%%%%%%%%%%%%%%%%%%%%%%%%%%%%%

In Fig.~\ref{fig:IPRH4}, we show {$ \overline{{\cal I}}$}  computed for the eigenstates of the static effective Hamiltonian $\hat{H}^{(4)}_{\rm eff}$  in comparison with the Floquet states obtained from ${\cal H}(t)$.  The results are qualitatively similar to those seen for $\hat{H}^{(2)}_{\rm eff}$ in Fig.~\ref{fig:2panels-2}, but there are quantitative differences. To illustrate  these differences, we mark in Fig.~\ref{fig:IPRH4} the place where, for small values of $g_3$, we get {$ \overline{{\cal I}} \approx 0.5$} for the second-order  static effective Hamiltonian $\hat{H}^{(2)}_{\rm eff}$ (gray circles), the fourth-order $\hat{H}^{(4)}_{\rm eff}$ (red squares), and the sixth-order $\hat{H}^{(6)}_{\rm eff}$ sixth order (white triangles). We see that the region of disagreement between Floquet states and eigenstates (blue region) decreases as the order increases.

In Fig.~\ref{fig:order}, we show the average IPR, {$ \overline{{\cal I}}$}, as a function of $g_3$, for a fixed value $g_4=10^{-7}$ (note that for large $g_3$, {$ \overline{{\cal I}}$} does not depend on $\g_4$),  for $\hat{H}_{\rm eff}^{(2)}$, $\hat{H}_{\rm eff}^{(4)}$, and $\hat{H}_{\rm eff}^{(6)}$. The horizontal line marks the point (circle) in the curves where {$ \overline{{\cal I}} \approx 0.5$}. One sees that the value of $g_3$ for {$ \overline{{\cal I}}\approx 0.5$} gets displaced to the right, effectively enlarging the area where there is good agreement between the eigenstates and Floquet states below the ESQPT. However, the rate of convergence of this expansion is slow and comes at the cost of highly complex expressions.

%%%%%%%%%%%%%%%%%%%%%%%%%%%%%%%
%%%%%%%%%%%%%%%%%%%%%%%%%%%%%%%%%%%%%%%%%%
%%%%%%%%%%%% FIGURE 8 %%%%%%%%%%%%%%
\begin{figure}
\includegraphics[width=.9\linewidth]{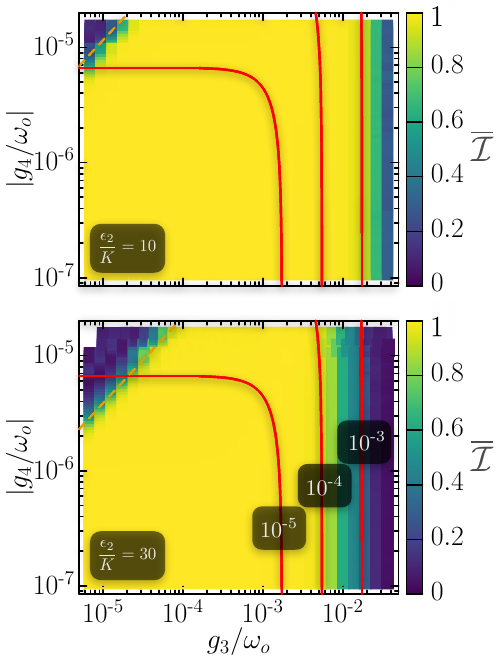}  %%{newfigIPRg4neg.png} 
\caption{\label{fig:g4neg} 
Average IPR,  {$ \overline{{\cal I}}$}, computed for the eigenstates of  $H_{\rm eff}^{(2)}$ in the Floquet basis as a function of $g_3/\omega_o$ and $|g_4/\omega_o|$, ($g_4<0$), for $\epsilon_2/K=10$ (top panel) and $\epsilon_2/K=30$ (bottom panel). Basis size $N=200$. The red lines in both panels are constant $K$ lines corresponding to $K/\omega_o=10^{-5},\,10^{-4}$, and  $10^{-3}$, as marked in the bottom panel. 
The dashed orange line in both panels corresponds to the empirically determined boundary $g_4=a g_3^{3/4}/(\epsilon_2/K)$ with $a=0.65$.
}
\end{figure}
%%%%%%%%%%%%%%%%%%%%%%

\section{Spectrum for negative $g_4$}
\label{app:negg4}
For completeness and because it can be important for present and future experiments,  we show  in Fig.~\ref{fig:g4neg}
the average IPR, equivalently to what was done in Fig.~\ref{fig:2panels-2} {for $H_{\rm eff}^{(2)}$}, but now for $g_4<0$. In this case, one cannot reach  $K=0$, but the results are very similar to those found for $g_4>0$.
%%%%%%%%%%%%%%%%%%%%%%%%%%%%%%%%%%%%%%%%%%%%%%%%%%%%%%%%%%%%%%
%%BIBLIOGRAPHY
\bibliographystyle{apsrev4-1} %{quantum}
\bibliography{bib_v2.bib}
%%%%%%%%%%%%%%%%%%%%%%%%%%%%%%%
\end{document}